\documentstyle[12pt]{article}
\input epsf

\def\bild#1#2{    
        \vspace*{-5mm}
        \begin{center}
        \begin{math}
        \epsfxsize#2cm
        \epsffile{#1}
        \end{math}
        \end{center}
        }

\begin{document}
\renewcommand{\thefootnote}{\fnsymbol{footnote}}
\begin{titlepage}
\renewcommand{\thefootnote}{\fnsymbol{footnote}}
\makebox[2cm]{}\\[-1in]
\begin{flushright}
\begin{tabular}{l}
TUM/T39-96-25
\end{tabular}
\end{flushright}
\vskip0.4cm
\begin{center}
{\Large\bf
Analytic Approximation to the GLAP evolution
of $F_2(x,Q^2)$ in the small-$x$ region. 
\footnote{Work supported in part by BMBF}}\\ 

\vspace{2cm}

L.\ Mankiewicz\footnote{On leave of absence from N. Copernicus
Astronomical Center, Polish Academy of Science, ul. Bartycka 18,
PL--00-716 Warsaw (Poland)} A. Saalfeld and T. Weigl 

\vspace{1.5cm}

{\em Institut f\"ur Theoretische Physik, TU M\"unchen, Germany}

\vspace{1cm}

{\em \today}

\vspace{1cm}

{\bf Abstract:\\[5pt]} \parbox[t]{\textwidth}{ We propose a systematic
  approximation scheme for solving GLAP evolution equations at small
  Bjorken-$x$. The approximate solutions interpolate smoothly between hard
  (singular as $x^{-(1+\lambda)}$, $\lambda> 0$) and soft (singular at most as
  $x^{-1}$) initial conditions and may be applied in a wide range of $Q^2$.
  The small-$x$ behavior of $F_2^p(x,Q^2)$  which is extracted from 
  a fit to HERA data agrees with results from a global fit.}

\vspace{1cm}
\end{center}
\end{titlepage}

\newpage Understanding the small-$x$ HERA data \cite{ZEUS,H1} 
is nowadays one of the most important challenges to perturbative QCD 
\cite{Theo96}. In a
conventional approach the link between theory and the data is provided by a
mathematical parameterization constructed to obey QCD evolution according to
Gribov-Lipatov-Altarelli-Parisi (GLAP) equations. Although global fits in the
whole $x$ region play necessarily the most important role
\cite{CTEQa,CTEQ4,GRV,MRS}, many approximations aimed specifically at the
small-$x$ region \cite{Forte95,Ral95,Kot95,Yndu96} have been proposed with a
goal to provide more insight \cite{Gehr96} into the otherwise numerically
complicated analysis.  As it has been emphasized in \cite{Weig96} such an
approximate procedure can be justified only if its accuracy with respect to the
exact calculation is higher than a typical accuracy of the experimental data.
In this letter we present an analysis based on a standard analytical approach
to the LO GLAP evolution equations which includes corrections necessary for
satisfactory agreement with the state-of-art numerical integration for both
hard, (singular as $x^{-(1+\lambda)}$, $\lambda> 0$) and soft (singular at most
as $x^{-1}$) initial conditions. By fitting the final formula to the data we
can explicitly demonstrate that the analytical approach predicts the same
small-$x$ shape of the input distributions as the global fit, as it should be.
Our analysis also shows that at $Q^2$ evolution lengths which are typical
for the present data, hard initial conditions may produce significant
contributions which possess a similar small-$x$ behavior as those resulting 
from soft initial conditions.  As a consequence, we have found that the
same data can be described by either moderately hard $\lambda \sim 0.2$ or soft
initial conditions, depending on the starting point of the evolution.

Let us now describe the analytic approach to the LO GLAP evolution which we
apply later to the analysis of the small-$x$ HERA data. As usual, we denote as
\begin{eqnarray}
q_s(n,t) & = & \int_0^1 dx\, x^n \, q_s(x,t)\, , 
\nonumber\\
g(n,t)   & = & \int_0^1 dx\, x^n \, g(x,t) \, 
\end{eqnarray}
the moments of the quark-singlet $q_s(x,t)$ and gluon $g(x,t)$ distributions at
a scale $t = \ln(Q^2/\Lambda^2_{QCD})$.  At leading order the GLAP evolution of
the quark singlet distribution in terms of the initial distributions
$q_s(n,t_0)$ and $g(n,t_0)$ has the well-known form \cite{QCDbook}
\begin{eqnarray}
q_s(n,t) &=& \left\{ (1 - h_2(n)) \, 
          q_s(n,t_0) - h_1(n) \, g(n,t_0) \right\} \,
          e^{\frac{2}{\beta_0} \Lambda_+(n)\zeta}
\nonumber \\
         &+& \left\{ h_2(n) \, q_s(n,t_0) + 
          h_1(n) \, g(n,t_0) \right\}
          e^{\frac{2}{\beta_0} \Lambda_-(n)\zeta }.
\label{mellinevolution}
\end{eqnarray}
In the above expression $\zeta=ln(\alpha(Q_0^2)/\alpha(Q^2))$ denotes the $Q^2$
evolution length and $\Lambda_{\pm}$ are the eigenvalues
of the anomalous dimension matrix.  The coefficients $h_1$ and $h_2$ may be
expressed through $\Lambda_{\pm}$ and the anomalous dimensions as
\begin{equation}
h_1(n) = \frac{\gamma_{qg}(n)}
         {\Lambda_-(n) - \Lambda_+(n)}
\; , \;\;\;
h_2(n) = \frac{\gamma_{qq}(n) - \Lambda_+(n)}
         {\Lambda_-(n) - \Lambda_+(n)},
\end{equation}
To obtain the Bjorken-$x$ distribution from (\ref{mellinevolution}) one has to
take the inverse Mellin transform which reads
\begin{equation}
x q_s(x,t) = \frac{1}{2\pi i} 
             \int_{c-i\infty}^{c+i\infty} dn\,
             x^{-n}\, q_s(n,t),
\label{mellininversion}
\end{equation}
where the integration contour runs to the right of all singularities of
$q_s(n,t)$. As we shall be ultimately interested in the small-$x$ behavior of
the quark distribution, let us assume that the initial conditions are given by
simple poles
\begin{equation}
q_s(n,t_0) = \frac{A_q(t_0)}{n-\lambda_q} 
\;,\;\;\;\;
g(n,t_0)   = \frac{A_g(t_0)}{n-\lambda_g}\;,
\label{singul}
\end{equation}
with $1 > \lambda_{q,g} \ge 0$.  Anticipating that such a power-like rise in
the initial condition is only valid for small enough $x$, one could introduce
an additional parameter $x_0$ above which one neglects the parton distribution
and below which one assumes a power-like behavior. With this interpretation in
mind it is evident that $0 < x_0 \le 1$.  This yields the following modified
ansatz for the initial boundary,
\begin{equation}
q_s(n,t_0) = \frac{A_q(t_0) \, x_{0,q}^{-\lambda_q} }
             {n-\lambda_q} \, x_{0,q}^n
\;,\;\;\;\;\;
g(n,t_0) = \frac{A_g(t_0) \, x_{0,g}^{-\lambda_g}}
             {n-\lambda_g} \, x_{0,g}^n,
\label{x0}
\end{equation}
where we have introduced separate $x_0$'s for quark and gluon distributions.
As far as the magnitude of $x_0$'s is concerned Ref.  \cite{Forte95}
convincingly argues that $x_{0,g}$ should be of order of $0.1$.
\\
We have found that the introduction of $x_{0 g,q}$ causes problems. From a
practical point of view the model parameters become highly correlated (cf.
equation (\ref{x0})). Secondly, the HERA data \cite{ZEUS,H1} favor $x_{0,q}$ to
be larger than one which contradicts the interpretation of $x_0$ as the
boundary of the power-like behaviour.  For these reasons we have decided to use
instead the following ansatz for the initial conditions
\begin{equation}
q_s(n,t_0) = \frac{A_q(t_0)}{n-\lambda_q} + B_q(t_0)
\;,\;\;\;\;\;
g(n,t_0) = \frac{A_g(t_0)}{n-\lambda_g} + B_g(t_0) \, .
\label{boundary}
\end{equation}
The constants $B_{q,g}$ model the leading corrections to a power-like small-$x$
boundary. Further corrections $n B_{g,q}^{(1)}(t_0) + \cdots $ can be
introduced if necessary.
\\
Our strategy for solving the Mellin inversion (\ref{mellininversion}) with
initial conditions (\ref{boundary}) will be the usual one, i.e., first we close
the contour in the left-half plane and then successively the residues are
taken, starting from those with the largest real parts.  The smaller $x$ is,
the less residues will be needed.  Therefore we first discuss the poles inside
the unit disc $|n|<1$. Apart from the pole at $n=\lambda$ the only other pole
singularity inside the unit disc occurs at $n=0$.  This can be read off from
the Laurent expansions of $\Lambda_{\pm}$ and $h_{1,2}$ which start like
\begin{eqnarray}
\frac{2}{\beta_0}\Lambda_+(n) &=& 
\frac{ {\tilde{a}}}{n} + \Lambda_+^R(n),
\nonumber \\
\Lambda_+^R(n) &=& - \delta_+ + {\cal{O}}(n),
\nonumber \\
\Lambda_-(n)  &=& {\cal{O}}(1), \nonumber \\
h_1(n), 1-h_2(n) &=& {\cal{O}}(n), 
\label{expansion1}
\end{eqnarray}
where ${\tilde{a}} = 12/\beta_0$ and $\delta_+=(11 + 2 n_f/27)/\beta_0$.  The
functions $\Lambda_+^R$, $\Lambda_-$ and $h_{1,2}$ are analytic inside the unit
disc.  From this one concludes that the contribution to the Mellin integral
(\ref{mellininversion}) from the $\Lambda_-$-term in (\ref{mellinevolution}) is
just the residue of the simple pole at $n=\lambda$
\begin{equation}
x q_-(x,Q^2) = 
A_q(t_0) \, h_2(\lambda_q) \, e^{a' \Lambda_-(\lambda_q) + b\lambda_q} +
A_g(t_0) \, h_1(\lambda_g) \, e^{a' \Lambda_-(\lambda_g) + b\lambda_g},
\label{polecontribution}
\end{equation}
where $a' = 2\zeta/\beta_0$ and $ b = \ln{(\frac{1}{x})}$.  The Mellin
inversion of the $\Lambda_+$-term in (\ref{mellinevolution}) is complicated
through the additional essential singularity at $n=0$.  Using that
$\Lambda_+^R$, and $h_{1,2}$ are analytic inside the unit disc one can formally
rewrite the Mellin inversion of the $\Lambda_+$-term as
\begin{eqnarray}
x q_+(x,Q^2) &=& 
A_q(t_0) \, {\hat U}_q(\partial_b,\zeta) \, J(a,b;\lambda_q) 
+ A_g(t_0)\, {\hat U}_g(\partial_b,\zeta) \, J(a,b;\lambda_g), 
\nonumber \\
 &+& \left\{ B_q(t_0) \, {\hat U}_q(\partial_b,\zeta)
     + B_g(t_0)\, {\hat U}_g(\partial_b,\zeta) \right\} \, 
     \left( \frac{a}{b} \right)^{\frac{1}{2}} \, I_1(\sqrt{2 ab}),
     \nonumber \\
J(a,b;\lambda) &=& \frac{1}{2 \pi i} 
\int_\gamma dn \frac{e^{{\frac{a}{n}+b n}}}{n-\lambda}
\label{Idefinition1}
\end{eqnarray}
where again $a = 12\, \zeta/\beta_0$, $ b = \ln{(\frac{1}{x})}$, $I_1$ is the
modified Bessel function \cite{Mathbook}, and $\gamma$ surrounds both poles
with positive orientation. The differential operators in (\ref{Idefinition1})
are constructed in such a way that their action reproduces the expansion
(\ref{expansion1}), e.g.,
\begin{equation}
{\hat U_g}(\partial_b,\zeta) = 
- h_1(\partial_b) e^{{\zeta \Lambda_+^{R}
(\partial_b)}} \, ,
\label{Udef}
\end{equation}
where $\partial_b = \frac{\partial}{\partial b}$ denotes the derivative over
the parameter $b$.  Note, that we do not expand the initial condition around $n
= 0$. Such an expansion would result in a term proportional to $1/\lambda$
which diverges at $\lambda = 0$.  It is clear that to obtain a sensible soft
initial condition limit one has to keep the initial condition term as a whole.

Now, from properties of modified Bessel functions it can be shown that
$J(a,b;\lambda)$ admits the following representation
\begin{equation}
J(a,b;\lambda) =  I_0(2 {\sqrt {ab}}) + 
\frac{\lambda}{2a} e^{b\lambda}   
\int_0^{{2 \sqrt {ab}}}  du \,u\, 
e^{-\frac{\lambda}{4a}u^2 } \, I_0(u) \,.
\label{Jreps}
\end{equation}
In the limit $\zeta=0=a$, i.e. no $Q^2$ evolution, the above formula exactly
reproduces the corresponding initial condition.
 
As the next step consider the operator ${\hat U}(\partial_b,\zeta)$. For fixed
$\zeta$ the function ${\hat U}(y,\zeta)$ should be looked at as a Taylor series
in $y$. After substituting $y$ by $\partial_b$, each term $y^i$ in the above
Taylor expansion becomes a differential operator that acts on $J(a,b;\lambda)$
to give
\begin{equation}
J_i(a,b;\lambda) = \frac{1}{2 \pi i} \, \int_\gamma dn \, n^i \,
\frac{e^{\frac{a}{n} +  b n}}{n-\lambda} \, . 
\label{iIdefinition}
\end{equation}
Again, the above integral can be evaluated to give
\begin{equation}
J_i(a,b;\lambda) = \lambda^i\sum_{k=0}^i 
\left(\frac{n_0}{\lambda}\right)^k
I_k(2 {\sqrt {ab}}) + 
\frac{\lambda^{i+1}}{2a}e^{b\lambda}   
\int_0^{2 {\sqrt {ab}}}  
du \,u\, e^{-\frac{\lambda}{4a} u^2} 
\, I_0(u)\, ,
\label{iIdefinition1}
\end{equation}
where $n_0 = \sqrt{\frac{a}{b}} $.
For a soft boundary $J_i(a,b;\lambda=0)$ simplifies to
\begin{equation}
J_i(a,b;\lambda=0) = n_0^i I_i(2 {\sqrt {ab}}).
\label{softJi}
\end{equation}
Combining the representation
\begin{equation}
I_k(z) = 
\frac{1}{\pi} \int_0^\pi dt 
\, e^{z \cos t} \cos{(kt)} 
\end{equation}
with the Taylor expansion of ${\hat U}(\partial_b,\zeta)$ it is possible first
to compute the geometric sums over $i$ in (\ref{iIdefinition}) and then to
evaluate their infinite series by a single complex integral. One obtains
\begin{eqnarray}
x q_+(x,t) 
&=&
A_q(t_0)\, \lambda_q \, U_q(\lambda_q,\zeta) \, \frac{e^{b\lambda_q}}{2a}
 \int_0^{2\sqrt{ab}} du \,u\, e^{- \frac{\lambda_q}{4a} \, u^2} \, I_0(u) 
\nonumber \\ 
&+& A_g(t_0)\, \lambda_g \, U_g(\lambda_g,\zeta) \,
\frac{e^{b\lambda_g}}{2a}  \int_0^{2\sqrt{ab}}  
du \,u\, e^{- \frac{\lambda_g}{4a} \, u^2} \, I_0(u)
\nonumber \\
&+& \frac{1}{\pi} 
\int_0^\pi dt \, e^{2 \sqrt{ab} \cos t} {\it{Re}}
\left( \frac{A_q(t_0) \, \lambda_q \, 
U_q(\lambda_q,\zeta)}{\lambda_q - n_0 e^{it}} 
      +\frac{A_g(t_0) \, \lambda_g \, 
U_g(\lambda_g,\zeta)}{\lambda_g - n_0 e^{it}}
\right) 
\nonumber \\
&-& \frac{n_0}{\pi}\int_0^\pi  dt 
\, e^{2 \sqrt{ab} \cos t}{\it{Re}}
\left(
\frac{ A_q(t_0) \, U_q(n_0 e^{it},\zeta) \,
e^{it}}{\lambda_q - n_0 e^{it}} + 
\frac{ A_g(t_0) \, U_g(n_0 e^{it},\zeta) \,
e^{it}}{\lambda_g - n_0 e^{it}}
\right) \, \nonumber \\
&+& \frac{n_0}{\pi} \int_0^\pi dt \, e^{2 \sqrt{ab} \cos t}
    {\it{Re}} \left\{ B_q(t_0) \, U_q(n_0 e^{it},\zeta)
    + B_g(t_0) \, U_g(n_0 e^{it},\zeta) \right\} \, ,
\label{long}
\end{eqnarray}
where ${\it Re}$ denotes the real part.  This formula is valid if $U(z,\zeta)$
admits a Taylor expansion around $z=0$ along the integration contours $z =
n_0\exp{(it)}$.  Thus, because the convergence radius of $U_{g,q}(z,\zeta)$ is
the unit disc, only data with $n_0<1$ should be compared with (\ref{long})
which is not a severe restriction in practice \cite{Weig96}.  Of course the
same restriction also applies to approximations that are constructed from
(\ref{iIdefinition1}), i.e. where only a finite number of terms are taken into
account.  The formula (\ref{long}) looks more complicated than it really is,
although it is necessary to take care about a proper definition of the
multi-valued complex square root used in the definition of the anomalous
dimension along the contour of integration.  In the soft boundary limit the
first two terms in (\ref{long}) vanish and the last term simplifies somewhat.
Apart from the kinematical restriction $n_0<1$, equation (\ref{long}) together
with (\ref{polecontribution}) yield the exact contribution to the parton
radiation arising from the leading singularity at $n=0$ of the anomalous
dimensions and the leading pole at $n=\lambda$ of the boundary.
\\
The next singularity in the Mellin plane at $n=-1$ can be treated along the
same lines that lead to (\ref{long}) but gives only a numerically small
correction at small $x$ because of an extra factor $x$ that multiplies most
terms.  An important difference now is the absence of a simple pole like the
one at $n=\lambda$ above. It can explicitly be shown that this causes all
contributions from the $n=-1$ singularity to vanish in the no-evolution limit
as it should, because the initial boundary is already recovered by the leading
poles at $n=\lambda$ and $n=0$.  Therefore one can expect that for moderate
$Q^2$ this term is additionally suppressed. In principle it is an easy task to
incorporate effects from the pole at $n=-1$ but for consistency one should then
also include the next pole from the initial conditions.  Rather than to proceed
this way we neglect all poles with $Re(n)\le -1$ and check the quality of the
approximation a posteriori by comparing with the exact result.  Figure 1 shows
such a comparison with the exact LO CTEQ \cite{CTEQ4} at $Q^2=10$\,GeV.  The
approximation (\ref{boundary}) to the initial conditions has been determined
from the CTEQ input parameterizations. For $q_-$ we use equation
(\ref{polecontribution}) and $q_+$ is constructed by taking in
(\ref{iIdefinition1}) the first non-vanishing term in the expansion of
$U_{q,g}(y,\zeta)$ around $y=0$.  One learns from this figure that when keeping
only data with, say, $x \le 5 \cdot 10^{-3}$, it is certainly justified to
neglect poles with ${\em Re} (n) \le -1$. Further, despite the fact that
typically $n_0$ is not particularly small \cite{Weig96}, it is not necessary to
evaluate the resummed formula (\ref{long}) but it is sufficient to take just
the first term from the expansions that are generated by (\ref{iIdefinition1}),
i.e.
\begin{eqnarray}
x q_s(x,t) & = & \left[ ( 0.198\, A_q(t_0) + 0.444\, A_g(t_0) ) \left( 
                 \frac{a}{b} \right)^\frac{1}{2} 
                 I_1(2 \sqrt{ab}) \right. \nonumber \\
           & + & ( 0.198\, \lambda_q \, A_q(t_0) 
                 + 0.444\, \lambda_g\, A_g(t_0) )\,
                 I_0(2 \sqrt{ab}) \nonumber \\
           & + & \left. ( 0.198 \, B_q(t_0) + 0.444 \, B_g(t_0) ) 
                 \left(\frac{a}{b} \right) I_2(2 \sqrt{ab}) \right] 
                 e^{-\delta_+ \zeta} \nonumber \\
           & + & \left[ 0.198 \, A_q(t_0) \, \frac{\lambda_q^2}{2 a} \,
                 x^{-\lambda_q} \, \int_0^{2 \sqrt{a b}} du\, u \, 
                 e^{- \frac{\lambda_q}{4 a} \, u^2}
                 I_0(u) \right. \nonumber \\
           & + & \left. 0.444 \, A_g(t_0) \, \frac{\lambda_g^2}{2 a}\,
                  x^{-\lambda_g} \, \int_0^{2 \sqrt{a b}} du \,u \, 
                 e^{- \frac{\lambda_q}{4 a} \, u^2}  
                 I_0(u) \right] e^{-\delta_+ \zeta} \nonumber \\
           & + & \left[ (1 - 0.198 \lambda_q)\, A_q(t_0)\, x^{-\lambda_q}
                 - 0.444\, \lambda_g\, A_g(t_0) \, x^{-\lambda_g} \right]
                 e^{-\frac{2}{\beta_0}\, 1.185 \,\zeta}.
\label{finform}
\end{eqnarray}
This is largely due to the fact that the Taylor coefficients that multiply
these terms are rather small for sub-leading terms and values of $\zeta$
typical for the current data. Note, that the extreme case where only the first
non-vanishing term in the expansion of $U_{q,g}$ is kept and
$\lambda_g,\lambda_q$ are set to zero corresponds to the soft boundary solution
that has been obtained in \cite{Forte95} by means of a saddle-point
approximation for large $\zeta$ and large $\xi$.

Furthermore we compare in Figure 1 our expansion scheme with a naive pole
approximation which takes into account the residue of the rightmost pole in
Mellin space at $n = \lambda_q, \lambda_g$ only.  A priori we expect such an
expansion scheme to work in the regime of large $\lambda$, where the pole is
sufficiently separated from the essential singularity at $n = 0$ arising from
the anomalous dimensions. Already for $\lambda$ values around $0.2 - 0.3$ as
represented by the CTEQ parameters (see Table \ref{table1}) the difference
between exact and approximated solution becomes larger than the experimental
errors. In case of yet smaller $\lambda$ we have checked that the discrepancy
strongly increases. Thus the naive pole approximation has to be replaced by the
more accurate formula, like equation (\ref{finform}), in a data analysis where
$\lambda$ is a variable fit parameter.

Using the above considerations we have performed a fit \cite{MINU} to the HERA
data \cite{ZEUS,H1} with the formula (\ref{finform}), starting from scales
$Q_0^2 = 1$ and $2.56$ GeV$^2$, the latter scale corresponding to the CTEQ
starting point. In both fits the number of quark flavors was globally assumed
to be $n_f=4$. The results are shown in Table \ref{table1}.  For orientation,
the table contains also the corresponding fit parameters determined from the
CTEQ initial conditions.  As far as the fit with the starting point $Q_0^2 =
2.56$ GeV$^2$ is considered, the parameters $A_q$ and $A_g$ as well as the
slopes $\lambda_g$ and $\lambda_q$ agree well with what has been found in
\cite{CTEQ4} from a global fit. It demonstrates the self-consistency of the
GLAP evolution in the small-$x$ region: the parameters of the fit which
determine the leading small-$x$ behavior are indeed determined mainly by the
small-$x$ data.  At the same time if one is interested only in the QCD
description of the small-$x$ data it is legitimate to use the analytical
approximation to the GLAP evolution, saving much effort as compared with a
global fitting procedure. The parameters $B_q$ and $B_g$ are reproduced less
accurately, but still the present fit seems to reproduce their correct signs.
Note, that $B_q$ comes out from both the CTEQ and our fits greater than zero
which contradicts the interpretation $x_{0,q}$, as we have discussed above.  
The relatively large errors are due to the fact 
that $B_q(t_0)$ and $B_g(t_0)$ generate non-leading contributions 
at small-$x$ and therefore are less sensitive to the data in this region.

The fit with a low starting point of $Q_0^2 = 1$ GeV$^2$ results with the same
data being parameterized by a soft gluon boundary, although the corresponding
$\lambda_g$ parameter is hardly constrained by the fit. It means that for a
sufficiently large evolution length soft and hard boundaries result in a
similar shape of $F_2(x,Q^2)$ at small-$x$.  Note, that although equation
(\ref{finform}) describes asymptotic behavior of the evolution starting from a
hard boundary, the first three terms on the right-hand side have a non
power-like, i.e. soft, small-$x$ behavior. To illustrate this point we show in
Figure 2 various contributions to $F_2(x,Q^2)$ for $Q^2 = 10$ and $1000$
GeV$^2$, respectively, as determined by our fit with the starting scale $Q_0^2
= 2.56$ GeV$^2$. As already stated in \cite{Weig96} it is important to include
$q_-(x,Q^2)$ in an analysis including data in a moderate $Q^2$ range. This
contribution is represented by the last term in Eq.(\ref{finform}).  For large
$Q^2$ this part of the solution gets strongly suppressed as expected from the
very beginning.  On the other hand the soft part, identified as the first three
terms on the right-hand side of equation (\ref{finform}), has a shape 
which is similar to the full solution. In other words, in the $x$ domain 
covered by the current
experiments soft and power-like initial conditions can be indistinguishable.
This observation may help to understand the known fact that the global fits can
explain the data equally well starting from soft \cite{GRV} or hard \cite{MRS}
initial conditions. 

Keeping $\lambda_g$ and $\lambda_q$ equal to zero, and allowing for the
starting point of the evolution to be as low as $0.45$ GeV$^2$, we have
obtained the fit presented in the last row in Table \ref{table1}. It is
consistent with results of \cite{GRV} where the LO fit starting from a yet
lower scale $0.24$ GeV$^2$ and valence-like distributions with $\lambda_g,\,
\lambda_q < 0$ was presented, but one should keep in mind that the
approximation (\ref{finform}) does not describe the GLAP evolution from such
initial conditions with a sufficient accuracy.

Finally we have investigated the double asymptotic scaling behavior of $F_2$
as function of the scaling variables $\rho = \sqrt{\xi /\zeta}$, $\sigma =
\sqrt{\xi \zeta}$ ($\xi = \ln(1/x)$, $\zeta =
\ln\frac{\alpha_s(Q_0^2)}{\alpha_s(Q^2)} )$, as advocated in \cite{Forte95}.
After rescaling $F_2$ by the factor
\begin{equation}
 R_F = N_F \, \rho \, \sqrt{\sigma} \exp(- 2 \, \gamma\, \sigma\, + 
       \delta_+ \, \sigma / \rho)
\label{dscale}
\end{equation}
where $\delta_+ = (11 + \frac{2}{27} n_f)/\beta_0$, we observe in Figure 3 that
the data accumulate in a flat area in the $\rho$-$\sigma$-plane, despite the
fact that the plot corresponds to a power-like initial quark distribution at
$Q_0^2 = 1$ GeV$^2$. In other words, the kinematic bounds of current
experiments restrict the position of the data points to a domain in the
$\rho$-$\sigma$-plane, in which $R_F F_2$ hardly differs from a flat plane. On
the other hand, the double-asymptotic scaling interpretation does not hold
anymore if the starting point is taken to be $Q_0^2 = 2.56$ GeV$^2$
\cite{For95}, mainly due to the redefinition of the $Q^2$ evolution length
$\zeta$, see Figure 4.

In summary, we have presented an approximation scheme for solving analytically
the GLAP evolution at small $x$ that treats corrections to the strict small-$x$
limit in a systematic way. In contrast to saddle-point approximations there is
no need to have $Q^2$ very large. Our solution interpolates smoothly between
soft and hard boundaries.  For moderately hard boundaries we have found that
the pole approximation alone is not sufficient to adequately reproduce the full
QCD evolution. The fit shows that it is not necessary to perform a global
analysis in the full Bjorken-$x$ range if one is interested only in the
parameters which determine the small-$x$ behavior of parton distributions at
the initial scale. We have confirmed the wisdom arising from the global fits,
that the slopes of initial quark and gluon distributions depend critically on
the starting point of the evolution. The same HERA data can be described by
either a hard boundary at $Q^2 = 2.56$ GeV$^2$ or a soft one if the starting
point is lowered below $Q^2 = 1.0$ GeV$^2$. The interpretation of the current
data in terms of double asymptotic scaling crucially depends on the magnitude
of the starting point of the $Q^2$ evolution.

\begin{table}[t]
{\hfill
\begin{tabular}{|l|c|c|c|c|c|} \hline
Data range& $Q_0^2$  & $A_g(t_0)/A_q(t_0)$      
& $B_g(t_0)/B_q(t_0)$  & $\lambda_g / \lambda_q$ & ${\chi}^2$/d.o.f. \\ \hline

$x < 0.005,$ & $2.56$ GeV$^2$& 0.84$\pm$0.08, & -0.81$\pm$1.02, & 0.27$\pm$0.02, & 0.78  \\ 
$Q^2 > 3{\rm GeV}^2$ & (fixed)    & 0.57$\pm$0.03  & 1.21 $\pm$2.33  & 0.18$\pm$0.01  &   \\ \hline

CTEQ \cite{CTEQ4}       
& $2.56$ GeV$^2$& 0.854, &-2.693, & 0.305, & \\ 
& (fixed)      &  0.502 & 1.799  & 0.200  &  \\ \hline

$x < 0.005,$         
& $1.0$ GeV$^2$& 0.19$\pm$0.05, & -1.62$\pm$0.49, 
& 0.32$\pm$0.03,  & 0.78  \\ 
$Q^2 > 3{\rm GeV}^2$ & (fixed)        
& 0.70$\pm$0.03  &  6.27 $\pm$0.77  
& 0.11$\pm$0.01 & \\ \hline

$x < 0.005,$         
& $0.45$ $\pm$ 0.01 GeV$^2$& 0.00$\pm$0.06, & 0.00$\pm$0.17, 
& 0.00 (fixed),  & 0.87  \\ 
$Q^2 > 3{\rm GeV}^2$ &        
& 1.04$\pm$0.04  &  1.61 $\pm$0.29  
& 0.00 (fixed) & \\ \hline
\end{tabular}
\hfill}
\caption[results]{Results of the fit of equation ({\ref{finform}}) to 
the small-$x$ HERA
data \cite{ZEUS,H1}. For orientation, the table contains also the corresponding
fit parameters determined from the results of the CTEQ LO global fit
\cite{CTEQ4}. In the last row we have kept $\lambda_g$ and $\lambda_q$ equal
to zero and  allowed for perhaps unphysically low starting point $Q_0^2$ in
order to present a fit with soft initial conditions for both quark and gluon
distributions. All errors are parabolic, see \cite{MINU} for details.}
\label{table1}
\end{table}

\pagebreak

{\bf Acknowledgments}\\
A.S. thanks Prof. J. Kwieci{\'n}ski for useful discussions. This work was
supported in part by BMBF, KBN grant 2~P03B~065~10 and German-Polish exchange
program X081.91.

\vfill\eject

\clearpage


\clearpage


\epsfbox{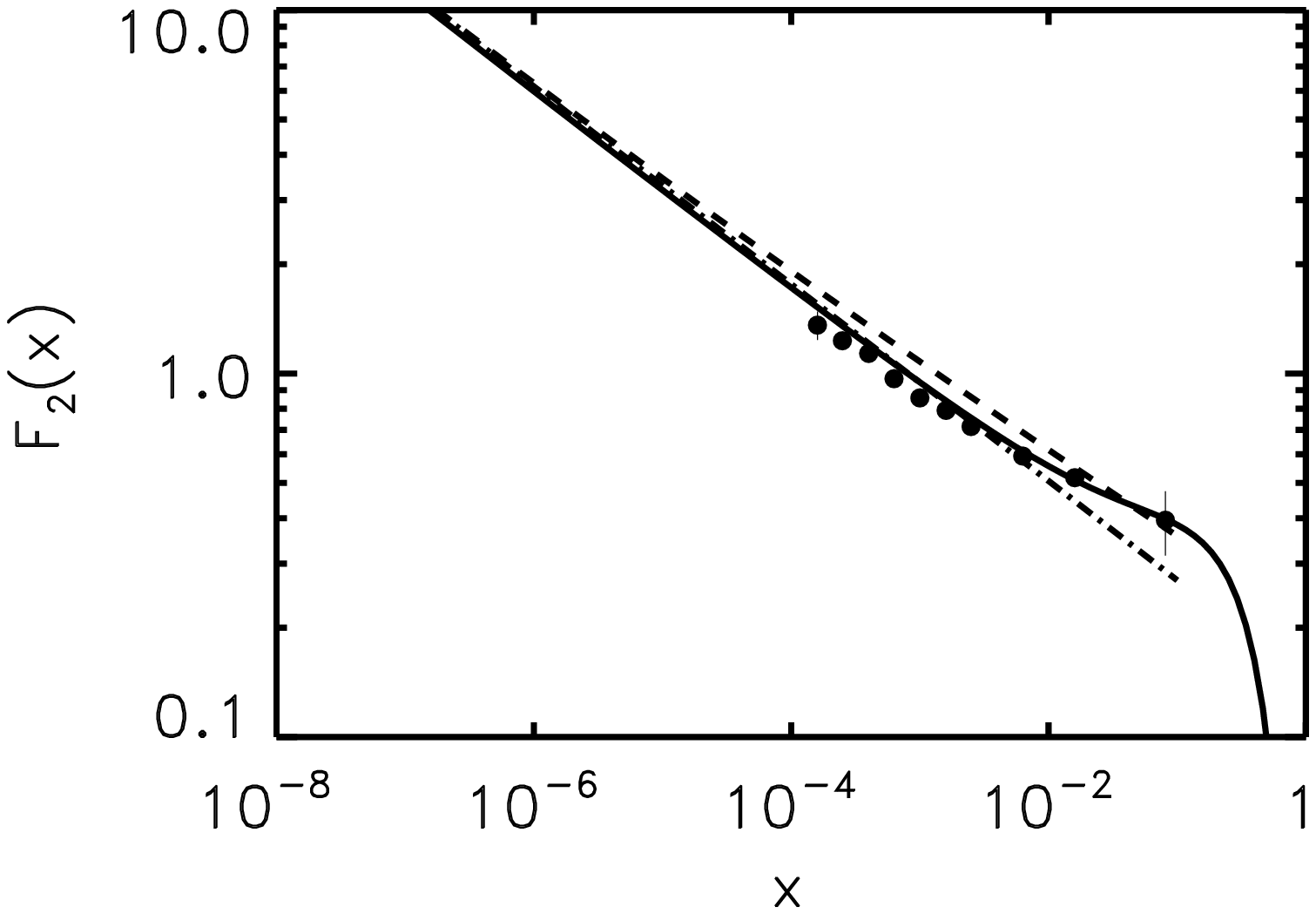}
\begin{description}
\label{figure1}
  
\item[Fig.~1] Comparison of the full (solid line) and the approximate evolution
  according to equation (\ref{finform}) (dot-dashed line) at $Q^2 = 10$
  GeV$^2$.  The initial conditions at $Q^2 = 2.56$ GeV$^2$ correspond to the
  CTEQ4L fit \cite{CTEQ4}.  The dashed line shows the result of the naive pole
  approximation, based on taking into account the residue of the rightmost pole
  in the Mellin plane only. Data points \cite{ZEUS} correspond to $Q^2 = 10$
  GeV$^2$.

\end{description}

\clearpage

\begin{minipage}{13cm}
\bild{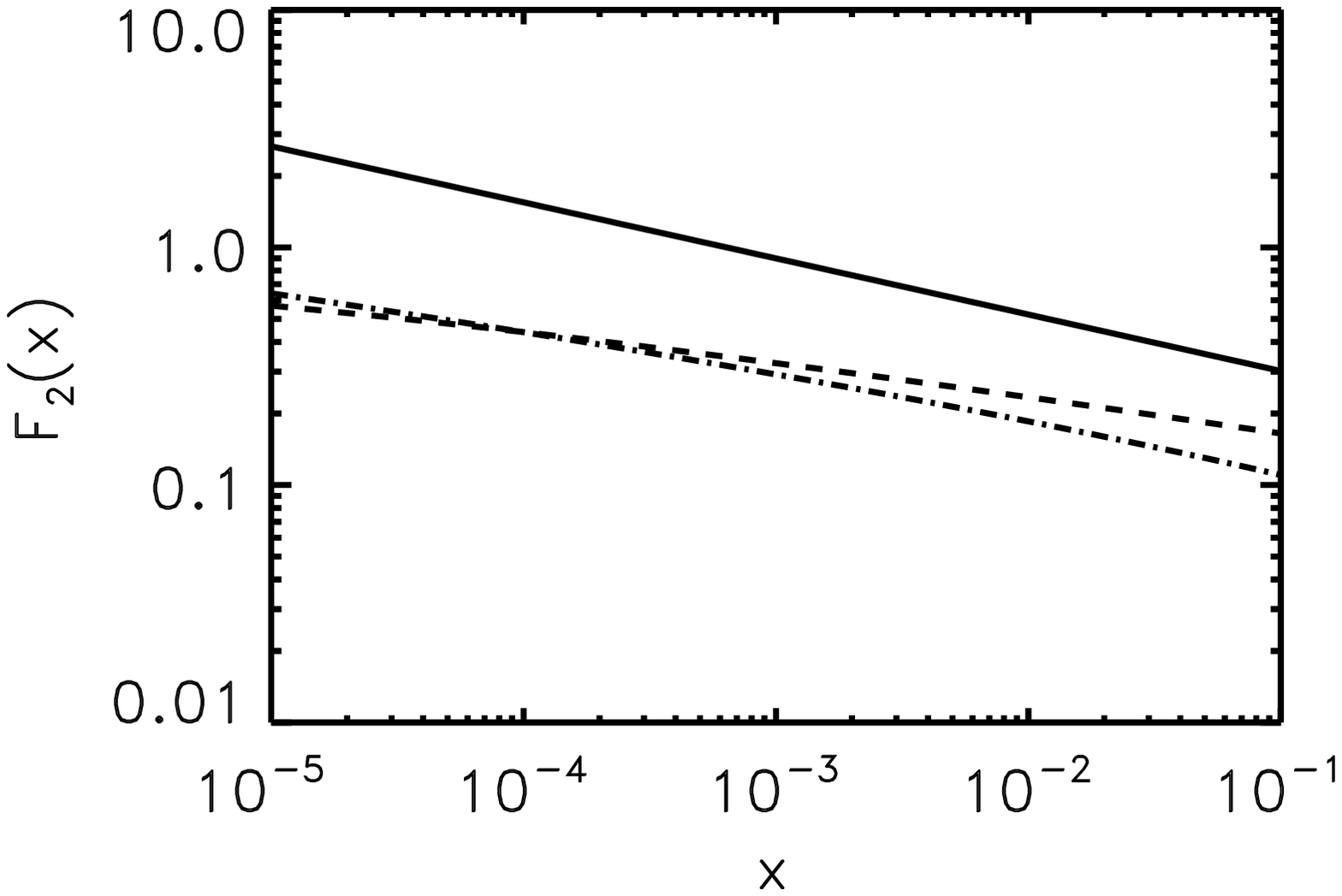}{13}
\bild{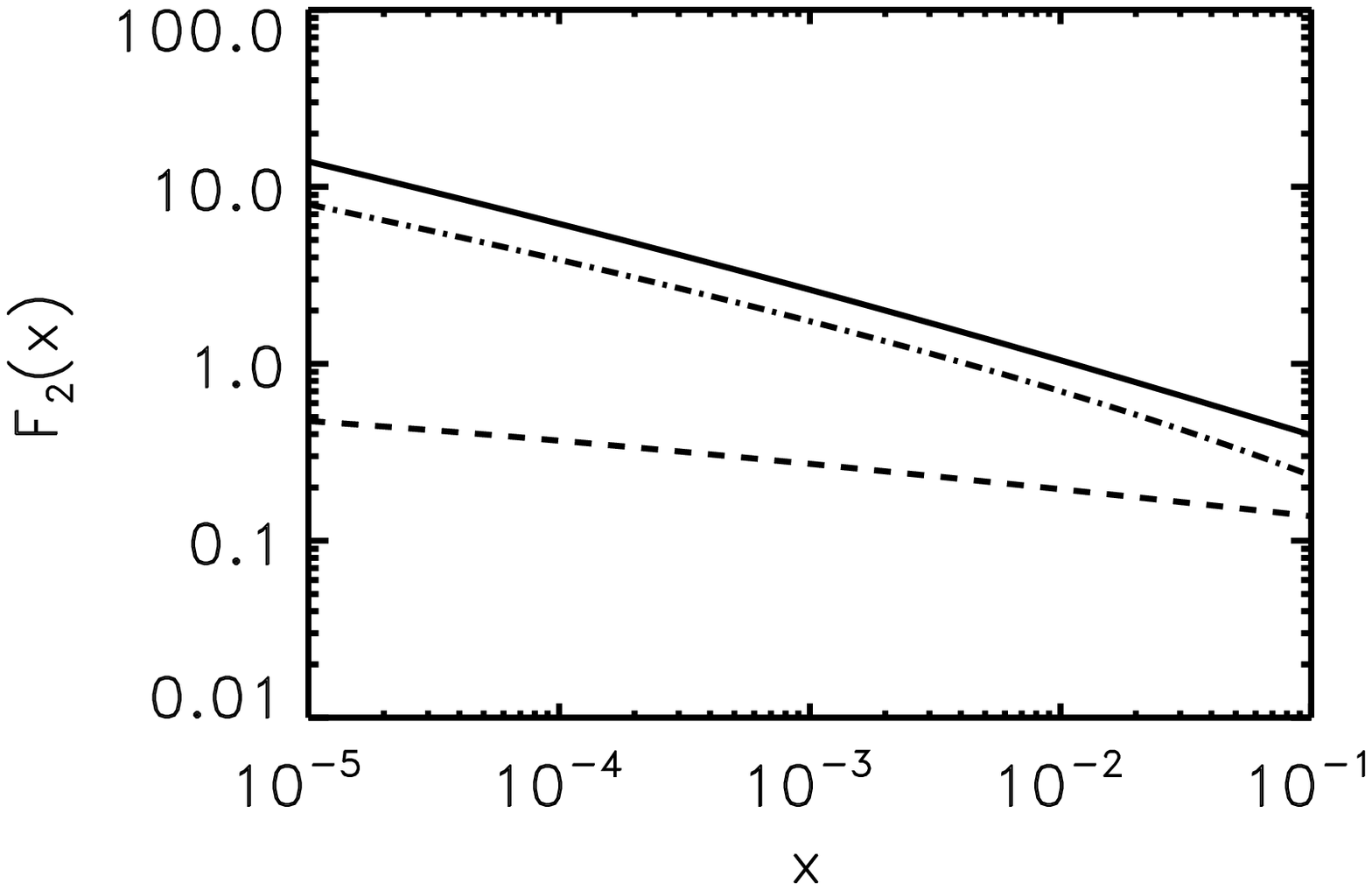}{13}
\end{minipage}

\begin{description}
 
\item[Fig.~2] Contributions of soft terms, and small eigenvalue $5/18
  q_-(x,Q^2)$ terms to $F_2(x,Q^2)$ at $Q^2 = 10$ GeV$^2$ (upper plot) and $Q^2
  = 1000$ GeV$^2$ (lower plot), as obtained from our fit with the starting
  scale $Q_0^2 = 2.56$ GeV$^2$. Full lines represent $F_2(x,Q^2)$.  Dot-dashed
  lines show the soft contribution, as given by the first three terms on the
  right-hand side of equation (\ref{finform}). Dashed lines show the small
  eigenvalue contribution $5/18 q_-(x,Q^2)$.

\end{description}

\begin{minipage}[t]{14cm}
\bild{figure3.eps}{13}
\end{minipage}

\begin{description}
\label{figure3}
  
\item[Fig.~3] $R_F F_2$ as a function of the scaling variables $\rho$ and
  $\sigma$ as given by the fit with $Q_0^2 = 1.0$ GeV$^2$. Dots denote 
  data points \cite{ZEUS,H1} with $Q^2 \ge 3$ GeV$^2$. Note that the 
  small-$x$ region corresponds to large values of $\rho$. 

\end{description}

\begin{minipage}[t]{14cm}
\bild{figure4.eps}{13}
\end{minipage}

\begin{description}
\label{figure4}
  
\item[Fig.~4] $R_F F_2$ as a function of the scaling variables $\rho$ and
  $\sigma$ as given by the fit with $Q_0^2 = 2.56$ GeV$^2$. Dots again 
  denote data points \cite{ZEUS,H1} with $Q^2 \ge 3$ GeV$^2$. The  
  $R_F F_2(\rho,\sigma)$ surface is not flat and the double-asymptotic 
  scaling interpretation does not hold anymore.

\end{description}

\end{document}